\documentclass[aps,prd,superscriptaddress,showpacs,twocolumn]{revtex4-1}
\usepackage{graphicx}
\usepackage[caption=false]{subfig}
\usepackage{epstopdf}
\usepackage{amsmath}
\usepackage{amsfonts} 
\usepackage{amssymb}
\usepackage{latexsym}
\usepackage{hyperref}
\usepackage[english]{babel}
\usepackage[utf8]{inputenc}
\usepackage[colorinlistoftodos]{todonotes}
\usepackage{xcolor}
\usepackage{slashed}
\usepackage{feynmp}
\usepackage{bm}
\usepackage{bbold}
\usepackage{eufrak}

\def \pt{\partial}

\def \L{\mathcal{L}}

\begin{document}
\title{Constraining hidden photons via atomic force microscope measurements and the Plimpton-Lawton experiment}

\author{D. Kroff}\email{daniel.kroff@gmail.com}
\affiliation{Avenida Francisco Alves 197, Rio de Janeiro 21940-260, Brazil}

\author{P.C. Malta}\email{pedrocmalta@gmai.com}
\affiliation{Rua Antonio Vieira 34, Rio de Janeiro 22010-100, Brazil}

\begin{abstract}
Modifications to electrodynamics from physics beyond the Standard Model can be tested to a high accuracy. Here we use two setups to place bounds on hidden photons, an Abelian boson kinetically mixed with the photon. The first setup involves atomic force microscope measurements, originally designed to study the Casimir effect at sub-$\mu$m distances. The second setup consists of two concentric metal shells with the outer one exposed to a high voltage. By measuring the potential difference between the shells it is possible to test Coulomb's law. The limits obtained here cover regions already excluded, in particular by astrophysical observations, but provide a more direct, laboratory-based confirmation of these bounds.
\end{abstract}

\maketitle


\section{Introduction} \label{sec_intro}
\indent

The Standard Model (SM) unifies electromagnetism to the strong and weak forces under the gauge group $SU(3)_{\rm C} \otimes SU(2)_{\rm L} \otimes U(1)_{\rm Y}$. Its predictions have been verified with great accuracy -- prime examples are the determination of the anomalous magnetic moment of the electron and the discovery of the Higgs boson. Despite of its successes, the SM does not account for a range of observations, such as the mass of neutrinos~\cite{Bilenky} or the nature of dark matter and energy~\cite{DM}. For overviews, see e.g. Refs.~\cite{Ellis, Kasakov} and references therein.

Extensions of the SM, such as string theory, seek to remedy these issues by expanding the symmetry group and, an often encountered possibility, is the inclusion of an extra $U(1)_{\rm X}$~\cite{Okun, Holdon, Abel1, Abel2, Fayet1}. This Abelian sector would mix with the weak hypercharge $U(1)_{\rm Y}$, thus leading to a mixing with the electromagnetic $U(1)_{\rm EM}$ in the low-energy limit. The new spin-1 boson associated with $U(1)_{\rm X}$, $X$, is electrically neutral and does not couple to SM matter fields directly. This means that, apart from the mixing with the photon, $X$ remains invisible, being dubbed hidden, or dark, photon.

The Lagrangian for the photon and hidden photon is, in natural units,
\begin{eqnarray}
\L = -\frac{1}{4}F_{\mu\nu}^2 - \frac{1}{4}X_{\mu\nu}^2 + \frac{\chi}{2}X_{\mu\nu}F^{\mu\nu} + \frac{m_{\gamma\prime}^2}{2}X_{\mu}^2 \, , \label{eq_Langrange}
\end{eqnarray}
where $X_{\mu\nu} = \pt_\mu X_\nu - \pt_\nu X_\mu$ is the field-strength tensor of the hidden photon. The unusual mixing in Eq.~\eqref{eq_Langrange} can be removed by field re-definitions~\cite{JJ}. There are two possibilities: $A \rightarrow A - \chi X$ or $X \rightarrow X - \chi A$. In the first case, the hidden photon and the electromagnetic current will interact, so particles with an electric charge $eQ$ acquire a hidden charge $-e\chi Q$; this is the origin of the so-called minicharged particles. In the second, the mixing is completely transferred to the mass terms, inducing a photon-hidden photon oscillation. For a review, see Ref.~\cite{Fabbrichesi}.


As will be discussed in Sec.~\ref{sec_pots}, the presence of hidden photons would induce a Yukawa-like modification to Coulomb's law, thereby affecting the interaction between electrically charged objects. It is therefore possible to look for new physics by analysing deviations from the standard predictions of Maxwell's electromagnetism. Here we analyse two experimental setups originally designed to search for very different effects. The first consists of a metallized plate and a sphere assembled in an atomic force microscope (AFM) and precisely measured the Casimir force at distances around 100~nm~\cite{Mohideen1, Mohideen2}. The second setup, implemented in 1936 by Plimpton and Lawton~\cite{Plimpton}, was composed of two concentric thin metal shells whose potential difference was accurately measured. It was therefore conceived as a sensitive test of Coulomb's law.

\begin{figure}[t!]
\begin{minipage}[b]{1.\linewidth}
\includegraphics[width=\textwidth]{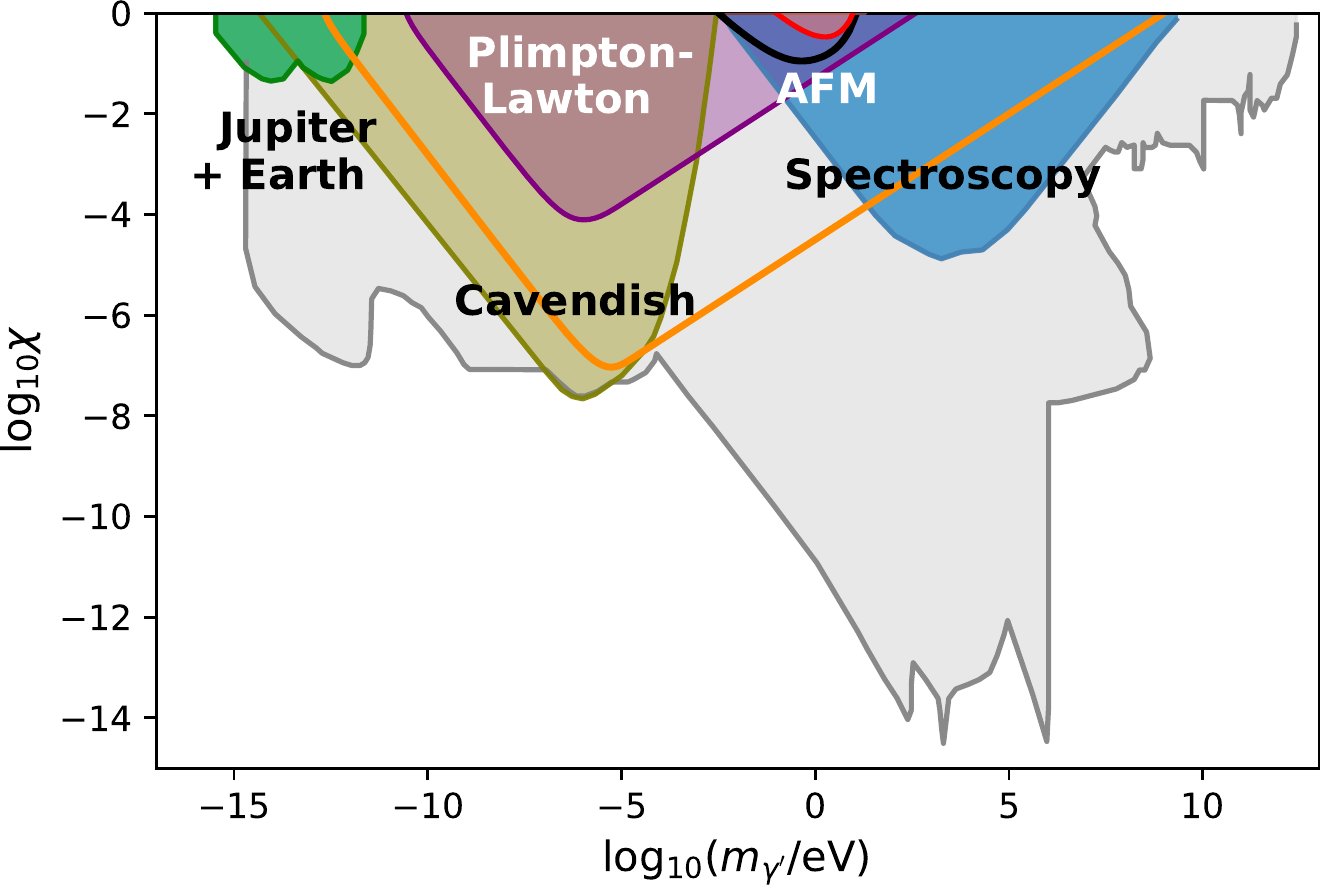}
\end{minipage} \hfill
\caption{Excluded regions in parameter space for hidden photons. Our bound from combining the force measurements from the AFM setups in Refs.~\cite{Mohideen1, Mohideen2} is displayed in red. In black we show the projected improvement for different sphere-plane separations and a 2-fold improvement in experimental uncertainty, cf.~Fig.~\ref{fig_exclusion_diff_a}. Our bound from the Plimpton-Lawton experiment~\cite{Plimpton} is shown in purple and, in orange, the projected improvement with better sensitivity and different radii, cf.~Fig.~\ref{fig_plimpton_alt}. Remaining contours adapted from Refs.~\cite{JJ, Caputo1, Caputo2}. For a discussion of hidden photons as dark matter candidate, see Ref.~\cite{Fabbrichesi} and references therein.}
\label{fig_exclusion_HP}
\end{figure}

The two AFM experiments considered here~\cite{Mohideen1, Mohideen2} are similar, consisting of a grounded gold-coated polystyrene sphere mounted on the tip of an AFM and placed at varying distances to an equally grounded gold-coated sapphire disk. Despite of the grounding, residual potentials of a few mV were measured between the sphere and the disk. These experiments are improvements upon previous runs~\cite{Mohideen3, Mohideen4} and display smaller experimental uncertainties -- around $1\%$ of the measured Casimir forces -- at the shorter disk-sphere separation distances. The reported rms theory-data deviations are of a few pN and indicate that theory and experiment are compatible at the $\sim 1 \, \sigma$ level. Departures from Coulomb's law must then lie under the experimental uncertainties and these results were used to exclude a region of the hidden-photon parameter space covering mass scales in the 0.1--10~eV range and $\chi \gtrsim 10^{-0.5}$,  as shown by the red curve in Fig.~\ref{fig_exclusion_HP}.


The second experiment considered was performed by Plimpton and Lawton~\cite{Plimpton}. Their setup consisted of two concentric spherical metal shells, the outer one being held at a harmonically varying potential. According to standard electromagnetic theory, in the interior of a charged conductor the electric field must vanish, i.e., the inner and outer shells should be at exactly the same potential.
The objective was to detect minute electric potentials induced in the inner shell, representing a precision test of Coulomb's law.
As discussed in Sec.~\ref{sec_plimpton}, the Yukawa-like term introduced by hidden photons would induce a non-zero electric field inside a conductor, thus leading to a potential difference between the shells. No such difference could be measured within the experimental uncertainties and this null result allows us to derive limits on the parameter space for hidden photons; the excluded region is displayed in purple in Fig.~\ref{fig_exclusion_HP}.

Incidentally, other Cavendish-like experiments involving multiple shells and precise measurements of potential differences between them have been performed to constrain the rest mass of the photon~\cite{Tu_2005}. These tests may be readily reinterpreted in terms of hidden photons and can be used to probe distance scales of order $\sim 10$~cm~\cite{Williams, Bartlett}. Atomic spectra would also be modified at first order and one can take advantage of the exquisite precision attained in measurements of e.g. the frequency in the $1s_{1/2}-2s_{1/2}$ transition in hydrogen to constrain hidden photons at atomic length scales~$\sim 0.1$~nm~\cite{Joerg_spectrum, Parthey}. Finally, static magnetic fields would also be affected and the non-observation of such effects at planetary scales has also been used to place constraints on very light hidden photons~\cite{Goldhaber1, Goldhaber2}. These limits are indicated as
``Cavendish", ``Spectroscopy" and ``Jupiter + Earth" in Fig.~\ref{fig_exclusion_HP}, respectively. Other limits from astrophysics, collider and laboratory tests are shown in gray. For further discussion on phenomenological limits, see Ref.~\cite{Ringwald}.

This paper is organized as follows: in Section~\ref{sec_pots} we derive the potential energy and the force between point charges due to the exchange of hidden photons. In Section~\ref{sec_limits_AFM} we analyse the AFM setups to determine the surface charge distributions on the conductors and calculate the force due to hidden-photon exchange, which is then compared to the experimental results. Next, in Section~\ref{sec_plimpton} we explore the results of the Plimpton-Lawton experiment and extract further limits. Finally, we dedicate Section~\ref{sec_conclusions} to our conclusions.


\section{Interaction potential mediated by hidden photons} \label{sec_pots}
\indent

Our target is to model the interaction between macroscopic bodies due to small modifications of electrostatics in the presence of hidden photons. In the AFM setups our observable is a force, whereas in the Plimpton-Lawton experiment we work with electric potentials. Both can be obtained once we know the potential energy between the objects involved, which can be traced back to the structure of the hidden photon propagator. This is an ultimately quantum mechanical calculation, but the potential energy is a classical quantity, so it is interesting to evaluate it in the non-relativistic (NR) limit. Here we assume static sources, thereby ignoring higher order velocity- and spin-dependent corrections~\cite{Holstein1, Holstein2, pots_HP_cbpf}.


From Eq.~\eqref{eq_Langrange} we see that hidden photons are, in principle, decoupled from the visible sector, with the exception of the mixing term, which may be treated as an interaction vertex $V^{\mu\nu}_{\gamma - \gamma^\prime} = i \chi \left( \eta^{\mu\nu}\, q^2 - q^\mu q^\nu \right)$, where $\eta_{\mu\nu}$ is the Minkowski metric and $q$ is the 4-momentum carried by the mediator. The sources are charged under $U(1)_{\rm EM}$, but not under $U(1)_{\rm X}$, so the effect of hidden photons in the interaction between them will be felt through the continuous oscillation of the mediating photons into hidden photons. It is expected that $\chi \lesssim 1$, what allows for a perturbative treatment of the corrections to the photon propagator. Inserting the mixing vertex between photon and hidden photon lines we obtain an effective propagator~\cite{pots_HP_cbpf}
\begin{eqnarray} \label{eff_prop}
\langle A_\mu A_\nu \rangle_{\rm eff} = -i\left( \frac{1}{q^2} + \frac{\chi^2}{q^2-m_{\gamma^\prime}^2}\right) \eta_{\mu\nu} + \mathcal{O}(\chi^4) \; ,
\end{eqnarray} 
where a gauge-fixing dependent term is omitted, since we are dealing with conserved external currents.


The potential energy between two elementary point charges separated by a distance $r = |{\bf r}|$ is given by the first Born approximation
\begin{equation}
U(r) = -  \int \frac{d^3 {\bf q}}{(2\pi)^3}  \, \mathcal{M}_{\rm NR} \,e^{i {\bf q}\cdot{\bf r}} \, , \label{born}
\end{equation}
with $\mathcal{M}_{\rm NR}$ being the NR limit of the relativistic amplitude $\mathcal{M}$. Here ${\bf q}$ is the 3-momentum exchanged between the sources in an elastic collision, in which $q^0 = 0$. The NR amplitude can be expressed as the contraction of the electrically charged source currents -- given by $J_i^0 \approx e$ and ${\bf J}_i \approx 0$ in the static limit -- with the effective propagator~\eqref{eff_prop}. The result, now expressed in SI units for later convenience, is~\cite{Joerg_spectrum, pots_HP_cbpf}
\begin{equation}\label{Energy_HD}
 U(r) = \frac{\alpha \hbar c}{r} \left(1 + \chi^2 e^{- m_{\gamma^\prime}c r/\hbar}\right)  \; ,
\end{equation}
where $\alpha = e^2/4\pi \epsilon_0 \hbar c$ is the electromagnetic fine-structure constant.


Equation~\eqref{Energy_HD} shows how the Coulomb potential is modified by a term screened by the mass of the hidden photon, i.e., a Yukawa-like term, which is further suppressed by the small mixing parameter $\chi$. This result is the basis for our analysis of the Plimpton-Lawton experiment in Sec.~\ref{sec_plimpton}. In order to incorporate the effects from hidden photons in the analysis of the AFM setups in Sec.~\ref{sec_limits_AFM}, it is useful to derive the electrostatic force between our point charges, which is given by
\begin{eqnarray} \label{eq_force}
F(r) = \frac{\alpha \hbar c}{r^2} \left[1 + \chi^2 \left(1 + \frac{m_{\gamma^\prime} c }{\hbar}r \right) e^{- m_{\gamma^\prime} c r/\hbar} \right] \; .
\end{eqnarray}

We note that for $\chi \rightarrow 0$ the usual inverse-square character of the Coulomb force is recovered. For $m_{\gamma^\prime} \rightarrow 0$, or rather for masses $m_{\gamma^\prime} \ll 1/d_{\rm exp}$, where $d_{\rm exp}$ is the typical length scale of the system under study, the hidden photon and the photon cannot be distinguished and the electromagnetic coupling constant is effectively redefined as $\alpha \rightarrow \alpha(1 + \chi^2)$. This observation is crucial when we try to extract meaningful bounds from the AFM setups in the small-mass region. For a very heavy hidden photon the exponential term is strongly suppressed and the hidden photon would not be excited at low energies, thereby leaving the Coulomb interaction unchanged.

Equations~\eqref{Energy_HD} and~\eqref{eq_force} display macroscopic effects that could give away the presence of hidden photons and are the basic results for the following analyses. The experiments considered here are sensitive probes to modifications of Maxwell's electromagnetism and can be used to constrain new physics covering distance scales ranging from $\sim 50$~nm to $\sim 10$~cm. Let us start with the AFM experiments.


\section{Limits from AFM measurements} \label{sec_limits_AFM}
\indent

The correction to Coulomb's force~\eqref{eq_force} is expected to be very small and only detectable at very short distances. Sensitive experiments are therefore necessary to search for it. Atomic force microscopes are particularly useful tools to measure very weak forces, such as the Casimir force -- the result of zero point vacuum fluctuations of the electromagnetic field~\cite{Casimir1, Lif}.

\begin{figure}[t!]
\begin{minipage}[b]{1.\linewidth}
\includegraphics[width=\textwidth]{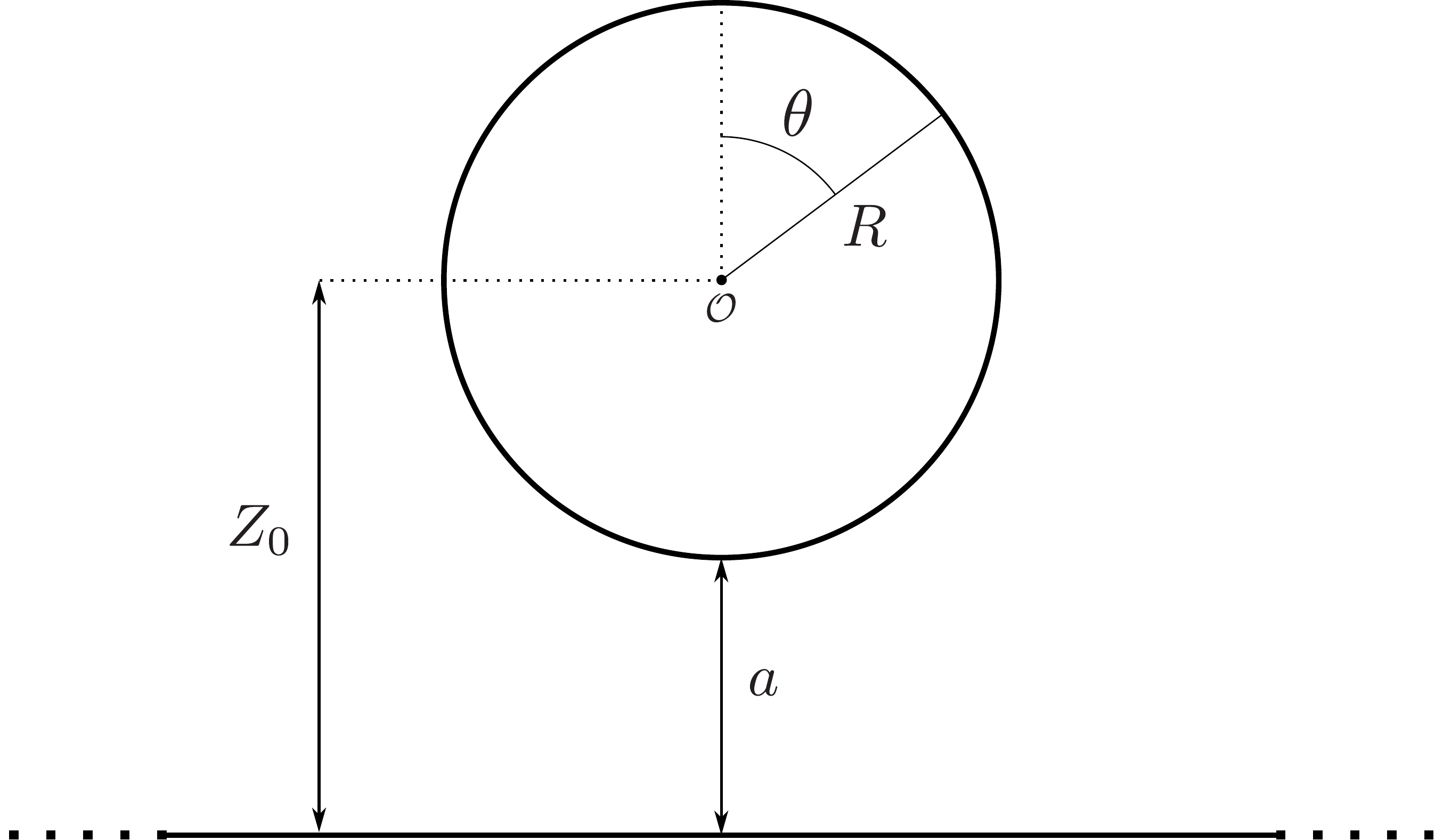}
\end{minipage} \hfill
\caption{The geometry of the AFM setups described in Refs.~\cite{Mohideen1, Mohideen2}. The point $\mathcal{O}$ -- the center of the sphere -- lies at a distance $Z_0 = R + a$ above the plane. The positions of the image charges, as well as their values, are functions of $Z_0/R$, cf. Sec.~\ref{sec_image}. Both sphere and plane are grounded, but there is a residual potential difference $V_0$ between them; see Table~\ref{table_AFM}.}
\label{fig_setup}
\end{figure}

The Casimir force has been demonstrated in a variety of geometries~\cite{Casimir2, Casimir3, Casimir4} and the agreement with theory -- including temperature effects and corrections for surface roughness and finite conductivity -- is in the 1$\%$ range. Due to the difficulty in arranging perfectly parallel plates at distances $\sim \mu$m, a commonly used geometry involves a sphere of radius $R$ at a distance $a \ll R$ from a plane, which is considered infinite in comparison with the other scales involved. This geometry is sketched in Fig.~\ref{fig_setup}.

The experiments considered here involve precise measurements of the force between a gold-coated polystyrene sphere of radius $R$ attached to the cantilever of an atomic force microscope and a similarly gold coated sapphire disk with 1~cm diameter~\cite{Mohideen1, Mohideen2}. Both the sphere and the disk are grounded, but residual potential differences $V_0$ of a few~mV between the objects could not be eliminated. The reported relative deviations from theory, i.e., Casimir and electrostatic forces, represent in both experiments $\sim 1\%$ of the measured forces, which are of the same order of magnitude as the experimental uncertainty. The relevant parameters are summarized in Table~\ref{table_AFM}. Therefore, the data are consistent with the theoretical predictions~\cite{Blocki, Mazur} and the results can only be used to set bounds on new physics. Other geometries have also been used to test new physics, such as non-Newtonian gravity~\cite{Novello1, Novello2, Bordag1} and other hypothetical particles~\cite{Bordag2}.

It is worth noting that a similar setup was investigated by Bordag {\it et al.}~\cite{Bordag3}. The authors considered a hypothetical Yukawa force felt by all atoms in each object and essentially integrated over the entirety of the respective volumes. Even though hidden photons also generate Yukawa-like forces, the analysis here is entirely different due to the coupling with electrically charged sources. Both polystyrene and sapphire are very poor conductors when compared to gold, so we may consider that there are no free charges in the bulks of the sphere and plate, which are taken to be electrically neutral as a whole. This means that the interior volumes of the polystyrene sphere and sapphire plate will not feel the Yukawa-like force~\eqref{eq_force}. The thin gold layers, on the other hand, are highly conducting and there is a small, but finite, residual potential difference between the objects driving the electric charges in the gold coatings to the surface, where they settle in non-trivial distributions. It is these induced superficial charges that will effectively contribute to the overall Yukawa-like force from hidden photons. In this sense, contrary to the analysis in Ref.~\cite{Bordag3}, the bulks do not contribute, whereas the charge distributions over the plane and the sphere are of utmost importance.


In the following we use the method of image charges applied to the AFM setups described in Refs.~\cite{Mohideen1, Mohideen2} to obtain the surface charge distributions on the plate and sphere and, by numerically integrating over these surfaces, we determine the expected force due to the exchange of hidden photons. 

\begin{table}
\begin{tabular}{|c|c|c|c|c|c|c|}
\hline
Ref. & $R$($\mu$m)  & $a$(nm)  & $V_0$(mV) & $\sigma_{\rm rms}$(pN) & $\sigma_{\rm exp}$(pN) & $\Delta M/M$ \\ \hline\hline
~\cite{Mohideen1} & 95.7 & 62 & 3 & 3.8 & 3.5 & $1\%$  \\ \hline
~\cite{Mohideen2} & 100.9 & 100 & 7.9 & 2.0 & 1.3 & $1\%$ \\ \hline
\end{tabular}
\caption{Basic parameters of the AFM setups considered.}\label{table_AFM}
\end{table}


\subsection{Electrostatic potential between perfectly conducting sphere and plane} \label{sec_image}
\indent

Our main goal is to use the reported results of AFM measurements~\cite{Mohideen1, Mohideen2} to place bounds on the parameters of a possible 
hidden photon modification to the Electrodynamics. Thus, we must, before anything else, understand how such a modification would affect the forces measured
in the experimental setup. Our strategy is simple: we obtain the charge distributions over both the sphere and the plane, which in turn are used to compute 
the Yukawa force between the two conductors.

Clearly, the hidden photon -- as well as any potential modification to the electromagnetic sector -- can only generate small deviation from Maxwellian
electrodynamics. In this spirit, we assume that the charge distributions over the conductors are sufficiently well described by the classical ones, even 
in the presence of the hidden photon. In this section we derive such distributions.

In line with the discussion above, we use two simplifying assumptions when modeling the AFM setup: we trade the disk for an infinite plane and treat both
the former and the sphere as perfectly conducting. Although apparently simple, we were unable to find, in the standard
electrodynamics textbooks (see, for instance Ref. \cite{jackson1999, *griffiths1999} and references therein), the solutions to the electric potential and 
fields for the configuration being considered; Refs. \cite{Morrison89, Dallagnol2009} were the only sources we came across that address this problem.

In Ref.~\cite{Morrison89}, the author approaches the problem through a power-series solution to Laplace's equation followed by a multipole expansion, in
a mathematical-physics \emph{tour de force}. The final result for the electric potential is given in terms of two power series: the first valid for the 
points whose distance to $\mathcal{O}$ is smaller than $2a$, and the second for points at distance greater than $2a$. However, the proposed solution fails 
for points exactly at a distance of $2a$ from $\mathcal{O}$. This is a major difficulty for our purposes, as we need to determine the surface charge density 
on the entire plane.

The solution proposed in Ref.~\cite{Dallagnol2009} is solely based on the method of image charges. This approach is not only very elegant and easy to follow, but
also yields expressions that are immediately suitable to our goals. Below we reproduce the main aspects of the argument and derive the expression that
will be used latter on.

It is well know that, given a set of boundary conditions, the solution to Laplace's equation is unique. Thus, if one manages to construct a solution 
fulfilling the boundary conditions, this must be the searched solution. The method of image charges explores this uniqueness in swapping the problem of 
interest by another one which is much simpler to analyse: a distribution of point charges. I.e., the idea is to place auxiliary point charges outside of 
the physical domain, in such a way that the boundary conditions of the original problem are satisfied. In our case, we shall place the image charges
inside the sphere and below the plane.

In order to apply the method of image charges, one must, first, determine the boundary conditions. As we are working in the perfect
conductor approximation, the sphere and the plane will represent equipotential surfaces. Thus, the boundary conditions will be simply given by the potential
at each of the conductors. As in the AFM setups there exists a residual potential $V_0$ difference between the probe and the disk, we can, with no loss of 
generality, demand that the potential on the plane vanishes, what makes the potential on the surface of the sphere $V_0$. 

Clearly, an image charge
\begin{equation}
Q_0 = 4\pi\epsilon_0V_0R,
\end{equation}
placed on the point $\mathcal{O}$, saturates the boundary condition on the surface of the sphere. Then, to ensure the condition $V=0$ on the plane, an image charge $-Q_0$ must be introduced. This charge must be positioned exactly below the original image charge and at a distance of $Z_0 = R + a$
below the plane.

Notice that the inclusion of the second image charge, while reinforcing the boundary condition over the plane, modifies the potential over the sphere. 
We can compensate for such a difference by including a third image charge, this time inside the sphere, but, then, a new one under the plane will also
be needed -- this procedure will lead to the inclusion of infinitely many pairs of image charges. 

In short, when a charge $Q_i$ is placed inside the sphere at a distance $Z_i$ from the plane, a companion charge $-Q_i$ must be placed exactly beneath it 
at a distance $Z_i$ below the plane. The image charges and their positions satisfy the following recurrence relations:
\begin{subequations}
\begin{equation}
    \zeta_i \equiv \frac{Z_i}{R} = \zeta_0 - \frac{1}{\zeta_0 + \zeta_{i-1}}
\end{equation}
\begin{equation}
    Q_i = \frac{Q_{i-1}}{\zeta_0 + \zeta_{i-1}}.
\end{equation}
\end{subequations}

It is interesting to notice that the position of the $N$th charge pair can be written in terms of a continued fraction as
\begin{equation*}
    \left.
    \begin{aligned}
    \zeta_N = \zeta_0 - \cfrac{1}{2\zeta_0 - \cfrac{1}{2\zeta_0 - \cfrac{1}{\ddots - \cfrac{1}{2\zeta_0}}}}
    \end{aligned}
    \right\} N \textrm{ denominators}.
\end{equation*}
Clearly this continued fraction converges for $N\to\infty$, meaning that, as $N$ increases, the image charges will be closer and closer to the 
accumulation point
\begin{equation}
    \zeta_{\infty} \equiv \lim_{N\to\infty} \zeta_N = \sqrt{\zeta^2_0 - 1}.
\end{equation}

\begin{figure}[t!]
\begin{minipage}[b]{1.\linewidth}
\includegraphics[width=\textwidth]{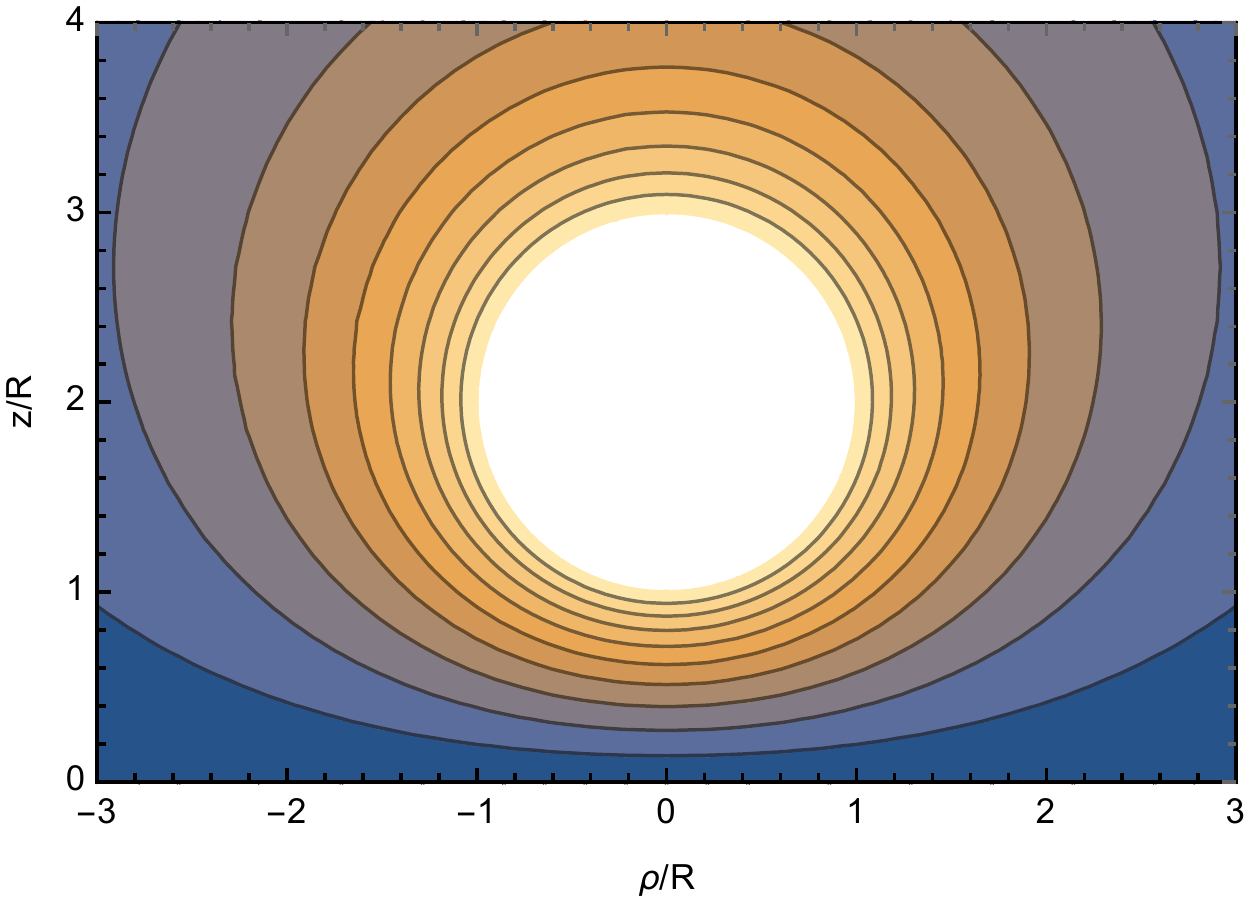}
\end{minipage} \hfill
\caption{Equipotential surfaces for the the AFM setups with a sphere of radius $R$ centered at $Z_0 = 2R$; the conducting plane, which is grounded, is located at $z = 0$. The white circle corresponds to the volume of the sphere, within which the potential is taken as constant. The boundary conditions at the surfaces of the conducting plane and sphere are duly fulfilled (color scale arbitrary).}
\label{fig_equipotentials}
\end{figure}

\begin{figure}[t!]
\begin{minipage}[b]{1.\linewidth}
\includegraphics[width=\textwidth]{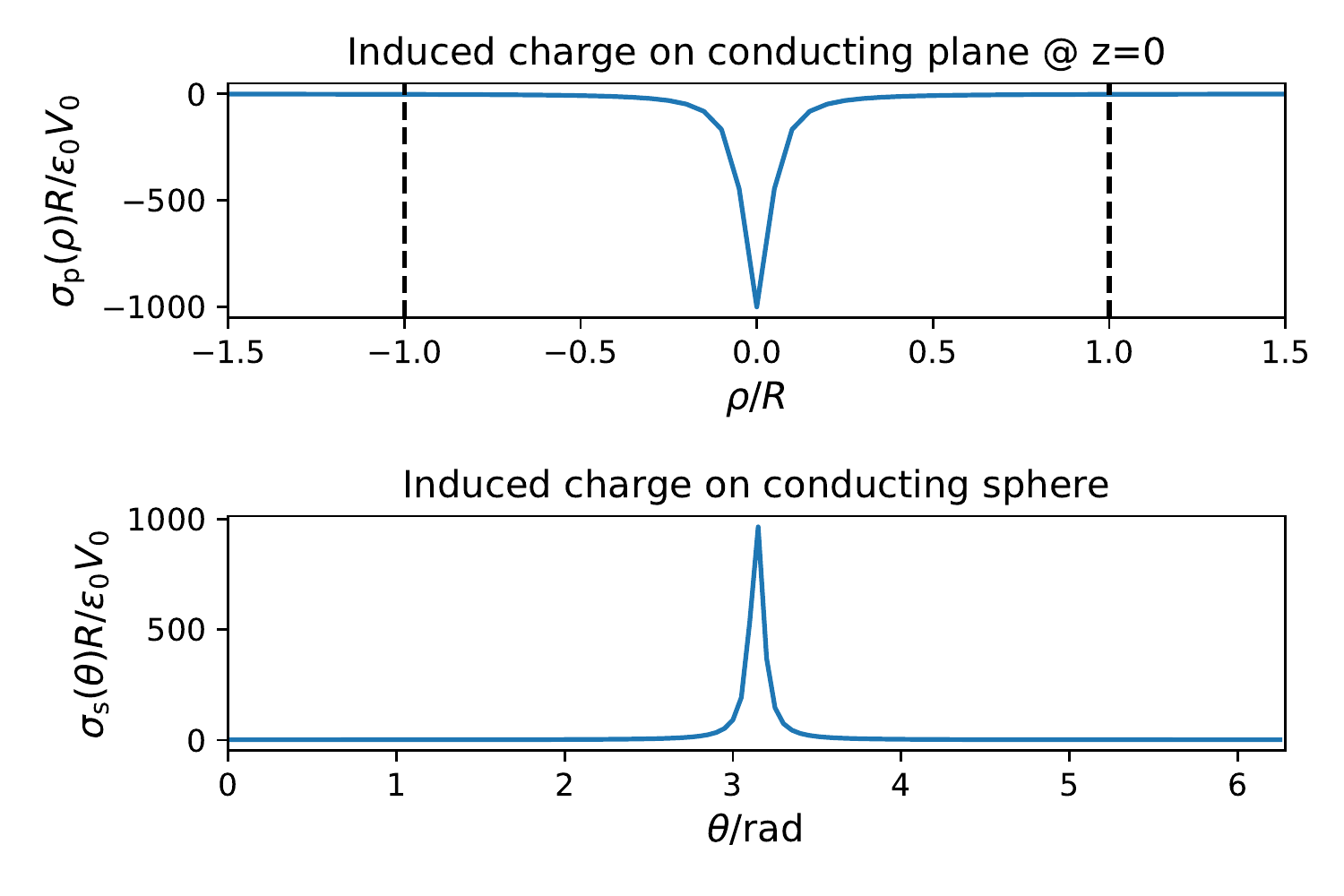}
\end{minipage} \hfill
\caption{Normalized surface charge densities for $Z_0/R = 1 + 0.001$. The dashed lines on the upper panel indicate the region on the plane directly below the sphere.}
\label{fig_charge_densities}
\end{figure}

Finally, for every point above the plane and on the outside of the sphere, the electric potential is given by the series
\begin{multline}
    V(z, \rho) = \frac{1}{4\pi\epsilon_0}\sum^\infty_{i=0}
    \left[\frac{Q_i}{\sqrt{(Z_i - z)^2 + \rho^2}}\right. \\
    \left. - \frac{Q_i}{\sqrt{(Z_i + z)^2 + \rho^2}} \right].
\end{multline}
The equipotential surfaces obtained from the expression above are shown in Fig.~\ref{fig_equipotentials} for $Z_0 = 2R$. The $z$ and $\rho$ components of the electric field in the region between the two conductors are given by
\begin{subequations}
\begin{multline}
    E_z(z,\rho) = -R\,V_0\sum^\infty_{i = 0}\left(
    \frac{Z_i - z}{[(Z_i - z)^2 + \rho^2]^{3/2}}\right. \\
    \left. + \frac{Z_i + z}{[(Z_i + z)^2 + \rho^2]^{3/2}} \right) \frac{Q_i}{Q_0},
\end{multline}
\begin{multline}
    E_{\rho}(z,\rho) = R\,V_0\sum^\infty_{i = 0}\left(
    \frac{\rho}{[(Z_i - z)^2 + \rho^2]^{3/2}}\right.\\ 
    \left.- \frac{\rho}{[(Z_i + z)^2 + \rho^2]^{3/2}} \right) \frac{Q_i}{Q_0}.
\end{multline}
\end{subequations}
From these expressions one can readily obtain $\sigma_{\rm p}$ and $\sigma_{\rm s}$, the surface charge densities over the plane and sphere (cf.~Fig.~\ref{fig_charge_densities}), respectively:
\begin{subequations}
\begin{equation} \label{eq_sigma_p}
    \sigma_{\rm p} (\rho) = \epsilon_0 E_z(0, \rho)
\end{equation}
\begin{equation} \label{eq_sigma_s}
    \sigma_{\rm s} (\theta) = \epsilon_0 \sqrt{E^2_z(z_{\rm s}, \rho_{\rm s}) + E^2_\rho(z_{\rm s}, \rho_{\rm s})},
\end{equation}
\end{subequations}
with $z_{\rm s} \equiv a + R(1 + \cos\theta)$ and $\rho_{\rm s} \equiv R\sin\theta$.

\subsection{Application to hidden photons} \label{sec_app_HP}
\indent



Due to the rotational symmetry around the z-axis, cf.~Fig.~\ref{fig_setup}, we may ignore any dependence on azimuthal or polar angles on the sphere and plane, respectively. This allows us to consider infinitesimal rings on the sphere and on the plane carrying infinitesimal charges $dq_{\rm s,p} = e \, dN_{\rm s,p}$ with
\begin{subequations}
\begin{eqnarray}
dN_{\rm s} & = & 2\pi R^2 \left( \frac{\epsilon_0 V_0}{e R} \right) \Tilde{\sigma}_{\rm s}(\theta) \sin\theta \, d\theta \label{dq_s} \\
dN_{\rm p} & = & 2\pi R^2 \left( \frac{\epsilon_0 V_0}{e R} \right) \Tilde{\sigma}_{\rm p}(\Tilde{\rho}) \, \Tilde{\rho} \, d\Tilde{\rho} \label{dq_p} \, ,
\end{eqnarray}
\end{subequations}
where $\Tilde{\rho} = \rho/R$. Here, $\Tilde{\sigma}_{\rm s}(\theta)$ and $\Tilde{\sigma}_{\rm p}(\Tilde{\rho})$ are the dimensionless surface charge distributions, cf.~Eqs.~\eqref{eq_sigma_p} and~\eqref{eq_sigma_s}.

The infinitesimal ring on the sphere is located at $z_{\rm s}$ and $\rho_{\rm s}$, whereas the infinitesimal ring on the plane is at $z_{\rm p} = 0$ and $\rho_{\rm p} = \rho$. The distance between them is given by $r^2 = (z_{\rm s} - z_{\rm p})^2 + (\rho_{\rm s} - \rho_{\rm p})^2$, which can be expressed as $r = R \Tilde{r}$, where
\begin{equation} \label{eq_dist}
\Tilde{r} = \sqrt{( \Tilde{a} + 1 + \cos\theta)^2 + (\Tilde{\rho} - \sin\theta)^2}
\end{equation}
with $\Tilde{a} = a/R$. Setting $\Tilde{m}_{\gamma^\prime} = m_{\gamma^\prime} cR/\hbar$ and using Eq.~\eqref{eq_force}, the infinitesimal vertical force between the two rings can be written as
\begin{eqnarray}
dF_{\rm z}^{\rm HP} & = & \alpha \chi^2 \hbar c \, \frac{(1 + \Tilde{m}_{\gamma^\prime} \Tilde{r}) e^{- \Tilde{m}_{\gamma^\prime} \Tilde{r}}}{r^2} \frac{z_{\rm s}}{r}  \, dN_{\rm s} \, dN_{\rm p}
\end{eqnarray}
so that, applying Eqs.~\eqref{dq_s} and~\eqref{dq_p}, we finally obtain
\begin{equation} \label{eq_force_HP}
F_{\rm z}^{\rm HP}(a) = \alpha \chi^2 \kappa \, I(a; m_{\gamma^\prime}) \, ,
\end{equation}
where $\kappa = \hbar c \left( 2\pi \epsilon_0 V_0 / e \right)^2$, which can be numerically expressed as
\begin{equation}
\kappa  =  3.8 \times 10^{-3} \, {\rm pN} \left( \frac{V_0}{{\rm mV}} \right)^2~
\end{equation}
and
\begin{equation}\label{eq_I}
I(a; m_{\gamma^\prime}) =  \int_{0}^{\infty} \! \int_{0}^{\pi} d\Tilde{\rho} \, d\theta \, \Tilde{\sigma}_{\rm s}(\theta) \Tilde{\sigma}_{\rm p}(\Tilde{\rho}) \, \mathcal{K}(\Tilde{\rho},\theta)  
\end{equation}
with
\begin{equation} \label{eq_K}
\mathcal{K}(\Tilde{\rho},\theta) = \frac{\Tilde{\rho} ( \Tilde{a} + 1 + \cos\theta) \sin\theta}{\Tilde{r}^3}  (1 + \Tilde{m}_{\gamma^\prime} \Tilde{r}) e^{- \Tilde{m}_{\gamma^\prime}\Tilde{r}} \, .
\end{equation}


The residual potential between the sphere and the plane generates surface charge distributions (cf.~Fig.~\ref{fig_charge_densities}) and Eq.~\eqref{eq_force_HP} is the hidden photon contribution to the total vertical electrostatic force between the objects. As the value of $\kappa$ indicates, this force is expected to be extremely weak. Incidentally, if we make $\chi = 1$ and $m_{\gamma^\prime} = 0$ in Eq.~\eqref{eq_force_HP} we recover the Coulombian electrostatic force. The function~\eqref{eq_K} encodes the mass dependence of the force and we expect it to be strongest at closest separation $d_{\rm exp} = a \sim \mathcal{O}(100 \, {\rm nm})$, i.e., for masses $m_{\gamma^\prime} \sim 1$~eV.

As already clear from Eq.~\eqref{Energy_HD}, for very small masses the Coulomb potential is re-obtained, but with $\alpha \rightarrow \alpha(1 + \chi^2)$. This means that, for very light hidden photons, the fine structure constant is substituted by the relation above, but this redefinition would not be observable, since $\alpha$ is experimentally determined. In other words, a massless hidden photon would be indistinguishable from the usual photon, so no modification to electromagnetic phenomena should be observed.

The arguments above imply that bounds on $\chi$ should weaken as we scan smaller masses. However, looking at Eqs.~\eqref{eq_force_HP} and~\eqref{eq_K}, we see that this is not the case: $F_{\rm z}^{\rm HP}$ tends to a nonzero, mass independent value. This is not physically meaningful. A similar situation is discussed in Ref.~\cite{Joerg_spectrum} in connection to the transition $1s_{1/2}-2s_{1/2}$ in hydrogen, for which the naive bound also fails to weaken as $m_{\gamma^\prime} \rightarrow 0$. The reason is the same: for small masses $\alpha$ is redefined and depends on $\chi$, i.e., we have two unknown parameters to be determined, $\alpha$ and $\chi$.

A solution to this problem is put forward by the authors of  Ref.~\cite{Joerg_spectrum}. They consider two independent measurements $M_i$ ($i = 1,2$), not necessarily of the same system, which are quoted in the form $M_i^{\rm exp}  - M_i^{\rm th} = \delta M_i + \Delta M_i$. Here $\delta M_i$ is the measured difference between theory and experiment, and $\Delta M_i$ is the experimental error. In our case the $\delta M_i$ are compatible with zero within the experimental errors for $\alpha = \alpha_0$ and $\chi^2 = 0$, i.e., in the absence of hidden photons. The $M_i^{\rm th}(\alpha, \chi^2)$ may then be expanded in a Taylor series around the unperturbed couplings $\alpha = \alpha_0$ and $\chi^2 = 0$ resulting in a system of linear equations in $\delta\alpha = \alpha - \alpha_0$ and $\chi^2$ that can be inverted to determine the values of the latter as

%
\begin{equation} \label{eq_chi2_1}
\chi^2 \leq \frac{ \left( \partial_\alpha M_1 \right)|\Delta M_2|  + \left( \partial_\alpha M_2 \right)|\Delta M_1|}{ \left( \partial_\alpha M_1 \right)\left( \partial_{\chi^2} M_2 \right) - \left( \partial_\alpha M_2 \right)\left( \partial_{\chi^2} M_1 \right)  }, 
\end{equation}
with the partial derivatives evaluated at $\alpha = \alpha_0$ and $\chi^2 = 0$. In more concrete terms, we may model the theoretical predictions as
\begin{equation} \label{eq_Mi}
M_i^{\rm th}(\alpha, \chi^2) = c_i \alpha^{n_i}\left[ 1 + \chi^2 f_i(m_{\gamma^\prime}) \right] \, ,
\end{equation}
where $c_i$ are dimensional factors and $f_i(m_{\gamma^\prime})$ contain the mass dependence of the observable. By plugging the equation above into Eq.~\eqref{eq_chi2_1} we finally obtain
\begin{equation} \label{eq_chi2_2}
\chi^2 \leq \frac{ \frac{n_1 |\Delta M_2|}{ M_2(\alpha_0, 0) } + \frac{n_2 |\Delta M_1|}{ M_1(\alpha_0, 0) }}{ |n_1 f_2(m_{\gamma^\prime}) -n_2 f_1(m_{\gamma^\prime}) |} \, .
\end{equation}

To see how this helps solving our problem, consider the massless case, where $M_i^{\rm th} = c_i \alpha^{n_i} \rightarrow c_i \alpha^{n_i} (1 + \chi^2)^{n_i}$. Given that $\chi^2 \ll 1$, we may expand $(1 + \chi^2)^{n_i} \approx 1 + n_i \chi^2$ and, comparing with Eq.~\eqref{eq_Mi}, we find that $f_i(m_{\gamma^\prime}) \rightarrow n_i$ for $m_{\gamma^\prime} \rightarrow 0$. Therefore, at this limit, the denominator of Eq.~\eqref{eq_chi2_2} tends to zero and $\chi$ increases, i.e., the bound weakens, as expected on physical grounds.

\begin{figure}[t!]
\begin{minipage}[b]{1.\linewidth}
\includegraphics[width=\textwidth]{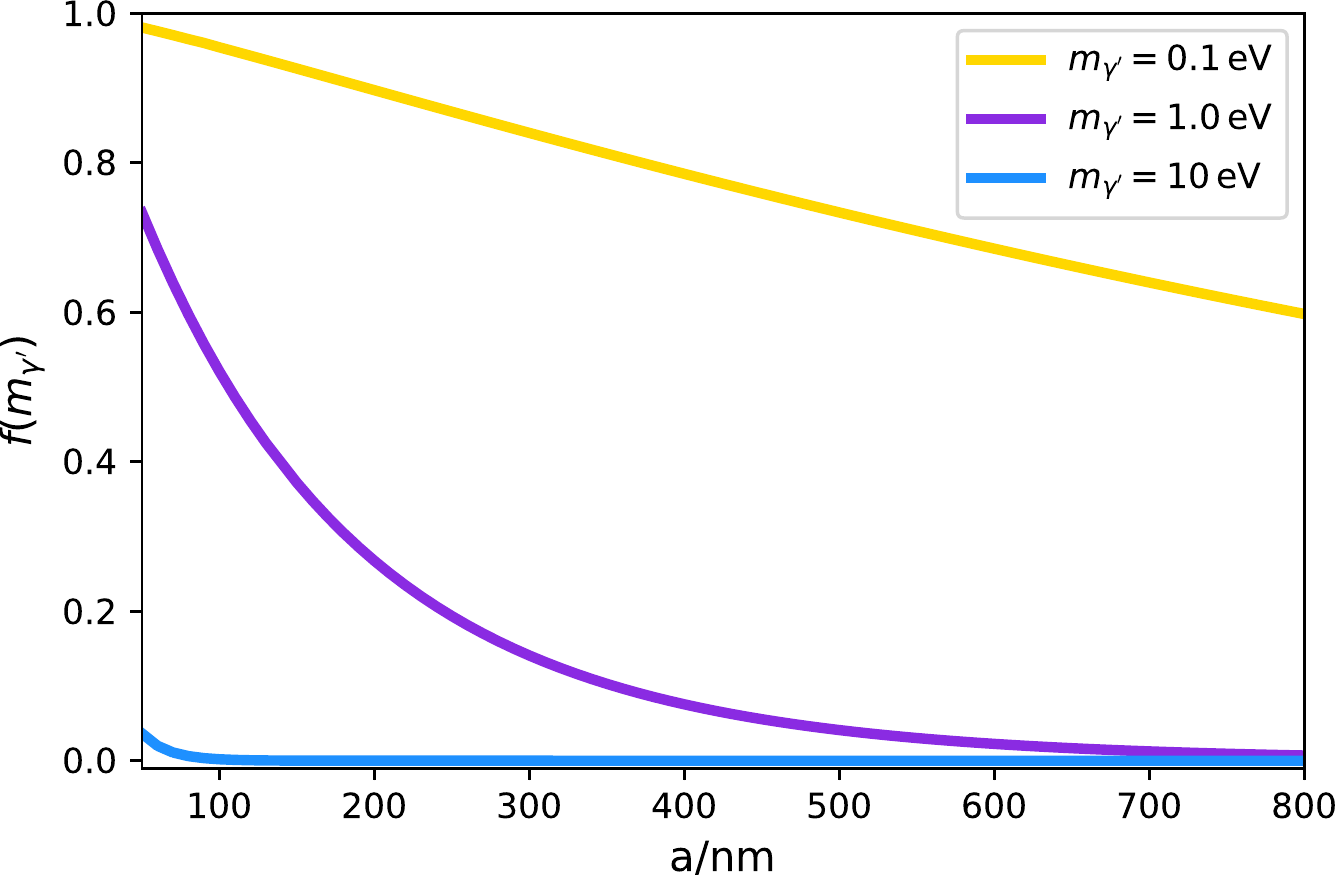}
\end{minipage} \hfill
\caption{The function $f(m_{\gamma^\prime})$, cf. Eq.~\eqref{eq_fi}, for different hidden photon masses as a function of the sphere-plane separation. It is, apart from a factor of $\chi^2$, the fractional deviation of the total electrostatic force due to hidden photons. For small masses, $f(m_{\gamma^\prime}) \sim 1$, which is a symptom of the ambiguity regarding the definition of $\alpha$. For heavy hidden photons $f(m_{\gamma^\prime}) \sim 0$, as it should.}
\label{fig_fm}
\end{figure}

We apply the procedure above by combining the AFM measurements in Ref.~\cite{Mohideen1} ($M_1$) and Ref.~\cite{Mohideen2} ($M_2$). From Eq.~\eqref{eq_force_HP}, we see that $n_1 = n_2 = 1$ and 
\begin{equation} \label{eq_fi}
f_i(m_{\gamma^\prime}) = I(a_i; m_{\gamma^\prime})/I(a_i; 0) \, .
\end{equation}
Incidentally, the $f_i(m_{\gamma^\prime})$ also represent the fractional deviation of the total force relative from the pure Coulombian contribution, modulo a factor of $\chi^2$. The dependence of $f_i(m_{\gamma^\prime})$ on the sphere-plane separation for different values of the hidden photon mass is shown in Fig.~\ref{fig_fm}.

\begin{figure}[t!]
\begin{minipage}[b]{1.\linewidth}
\includegraphics[width=\textwidth]{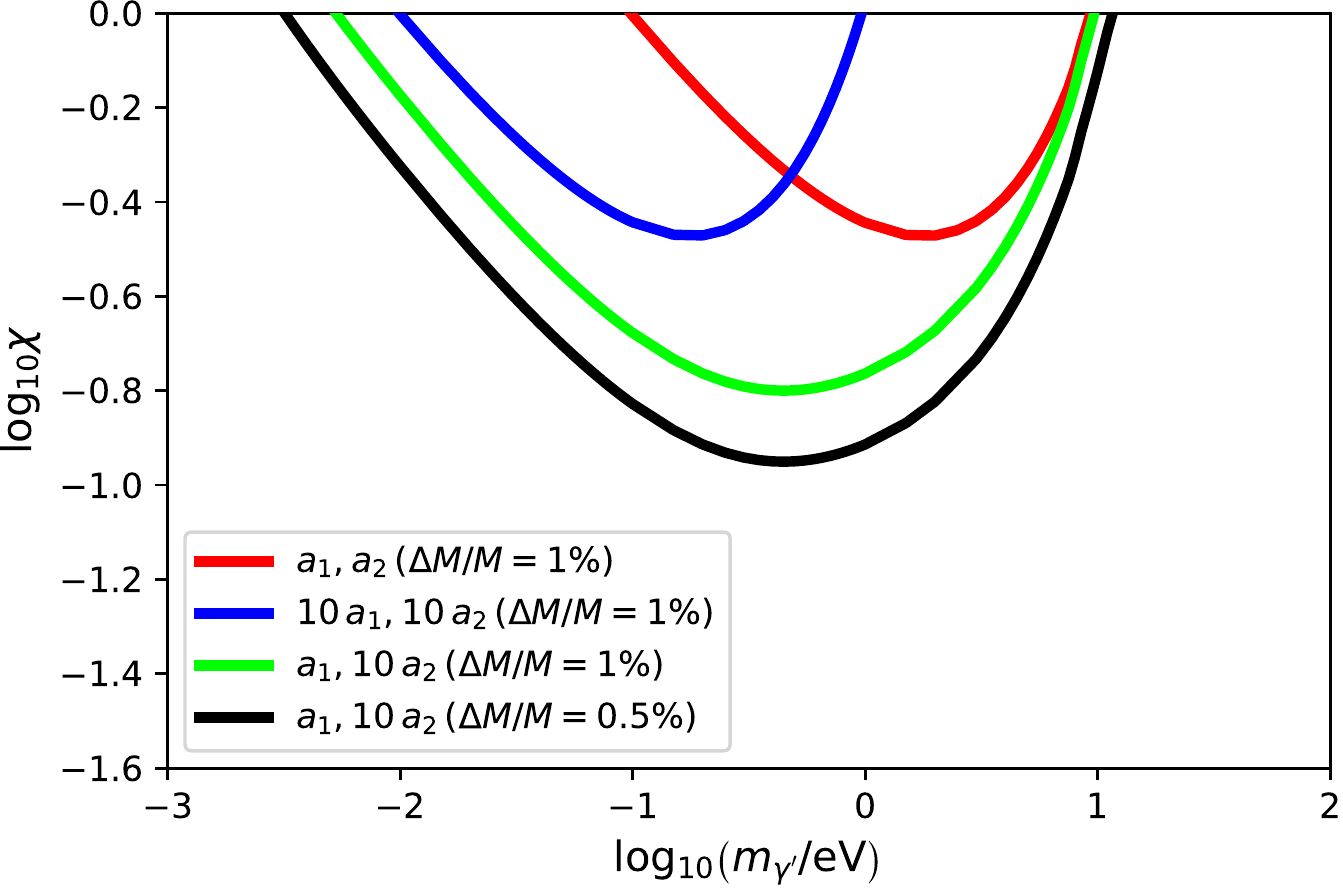}
\end{minipage} \hfill
\caption{Bounds on hidden photons from combining AFM data at different sphere-plane separations. Our bound with $a_1 = 62 \, {\rm nm}$ and $a_2 = 100 \, {\rm nm}$~\cite{Mohideen1, Mohideen2} is shown in red. In blue and green are the projected improvements when other separations are considered while keeping the relative experimental uncertainties at $1\%$ -- for the black curve we assume a realistic two-fold improvement to $0.5\%$.}
\label{fig_exclusion_diff_a}
\end{figure}

Using Eqs.~\eqref{eq_chi2_2} and~\eqref{eq_fi} with $|\Delta M_i|/M_i(\alpha_0, 0) = 10^{-2}$, cf.~Table~\ref{table_AFM}, we obtain the excluded regions shown in Fig.~\ref{fig_exclusion_diff_a}. In particular, the red curve shows the bound with the parameters reported in Refs.~\cite{Mohideen1, Mohideen2}, which exhibits the physically correct behavior for both small and large masses. This bound is also shown in Fig.~\ref{fig_exclusion_HP} in the context of previous limits; it is clear that it covers an area already excluded by atomic-physics tests~\cite{Joerg_spectrum}.


Given that we are not able to avoid other laboratory-based limits, it is interesting to discuss possible improvements that could be pursued in future AFM experiments. We expect the bound to be most sensitive for  $m_{\gamma^\prime} \sim 1/a$. This is clear from the red curve in Fig.~\ref{fig_exclusion_diff_a}, which displays our bound combining measurements at separations $a \sim 100$~nm. Note that $a_1$ and $a_2$ are relatively similar, so it is worthwhile checking how the bounds would behave for other choices of sphere-plane separation. To this end, in Fig.~\ref{fig_exclusion_diff_a} we also show the projected bounds for different combinations of sphere-plane separations $a_1 < a_2$. We see that, simultaneously increasing $a_1$ and $a_2$ by a common factor of ten, we obtain the blue curve, which is identical to the red one, but shifted to the left towards smaller masses. Had we reduced the separations, the bound would have moved to the right, though this direction is certainly more challenging to implement from a practical point of view.

Finally, it is clear that the extent of the high(low)-mass branch of the bound is determined by the shorter(larger) separation. The green curve in Fig.~\ref{fig_exclusion_diff_a} exemplifies a situation where $a_1$ is held fixed, but $a_2$ is increased, thereby not only improving the bound in terms of couplings excluded, but also extending the covered mass range. This result does not include a possible improvement on the experimental uncertainty, but, assuming the same parameters and only reducing the uncertainty by a modest factor of two, we obtain the black curve. This shows that simply increasing the sphere-plane separation can be a promising strategy as to directly probe, via  precise AFM force measurements, the area between the ``Cavendish" and ``Spectroscopy" limits in Fig.~\ref{fig_exclusion_HP}, which is already constrained by more indirect astrophysical tests.

\section{Limits from the Plimpton-Lawton experiment}\label{sec_plimpton}
\indent

Coulomb performed his famous experiments in 1785, but already in 1773 Cavendish tested the electromagnetic interaction between charges by using concentric metallic shells. His experiments were designed to show that free charges flow to the surface of conductors rather than accumulating in the bulk volume. The observed absence of charge in the interior of the conducting body could be explained if the electric field followed an inverse-square law. In fact, his null results were precise enough to be used to place limits on deviations from this behavior. A very important scenario where this could happen is the case of a small, though finite, photon mass. For an excellent review on this issue, see Ref.~\cite{Tu_2005}.

In 1936 Plimpton and Lawton~\cite{Plimpton} devised an experiment to test the validity of Coulomb's law by employing two concentric conducting spherical shells of radii $R^* = 0.61$~m and $R = 0.76$~m, a geometry close to that used by Cavendish in his original experiments~\cite{Tu_2005}. The internal shell is grounded, while the external one is subject to a
2~Hz harmonically-oscillating potential with and amplitude $V_0 = 3$~kV. According to standard electromagnetic theory, there should be no potential difference between the two shells. However, if the inverse-square law is not exact, this conclusion would not hold.

Let us now consider a conducting sphere  with an overall electric charge $Q$ in the presence of hidden photon. Due to the spherical symmetry of the problem, the electric charge will be homogeneously distributed on the surface and the potential is given by

\begin{eqnarray}
V(r) & = & \frac{Q}{8\pi\epsilon_0 Rr } \Bigg[ r + R - |r - R| \nonumber \\
&  & \quad \quad -\frac{\chi^2}{\Tilde{m}} \left( e^{-\Tilde{m}(r + R)} - e^{-\Tilde{m}|r - R|} \right)    \Bigg]  \label{eq_pot_plimpton}
\end{eqnarray}
with $r$ the distance to the center of the sphere and $\Tilde{m} = m_{\gamma^\prime}c/\hbar$. Note that, for $\chi = 0$, the potential is constant inside the sphere, as it should. However, if hidden photons are present and $\chi \neq 0$, there is a clear deviation from the constant potential due to the exponential terms, thus indicating a non-vanishing electric field in the interior of a conductor.

In the limit of infinitely heavy hidden photons we recover the usual Coulomb potential, as required. However, for very small masses, there is an effective redefinition of the interaction strength -- this is the same situation discussed in Sec.~\ref{sec_limits_AFM}. Even though this amounts to a renormalization of the coupling constant, the procedure given in Ref.~\cite{Joerg_spectrum} is not needed, as will be clear from the discussion below.

We are interested in the relative potential difference between the outer shell at $r = R$ and the inner shell at $r = R^* < R$, that is $\Delta V / V(R) \equiv \gamma$. Using Eq.~\eqref{eq_pot_plimpton} we obtain
\begin{equation} \label{eq_gamma}
\gamma \equiv 1 - \frac{V(R^*)}{V(R)} = 1-  \frac{ 1 + \chi^2 h(m_{\gamma^\prime}) }{ 1 + \chi^2 g(m_{\gamma^\prime}) } \, ,
\end{equation}
where
\begin{subequations}
\begin{eqnarray}
h(m_{\gamma^\prime}) & = & \frac{1}{2\Tilde{m}R^*} \left( e^{-\Tilde{m}(R - R^*)} - e^{-\Tilde{m}(R + R^*)} \right) \label{eq_h} \\
g(m_{\gamma^\prime}) & = & \frac{1}{2\Tilde{m}R} \left(1 - e^{-2\Tilde{m}R} \right) \, . \label{eq_g}
\end{eqnarray}
\end{subequations}
By manipulating Eq.~\eqref{eq_gamma} it is easy to write $\chi$ in terms of $m_{\gamma^\prime}$ as
\begin{equation} \label{eq_limit_plimpton}
\chi = \sqrt{ \frac{\gamma }{(1 - \gamma)g(m_{\gamma^\prime}) - h(m_{\gamma^\prime})  } } \, .
\end{equation}

It is worthwhile noting that, contrary to the potential~\eqref{eq_pot_plimpton}, Eq.~\eqref{eq_limit_plimpton} shows the correct behavior in both massless and massive limits, which are given by
\begin{subequations}
\begin{eqnarray}
\lim_{m_{\gamma^\prime} \rightarrow 0} \chi^2 & = & \frac{6\gamma \, \Tilde{m}^{-2}}{R^2 - R^{*2} - 6\gamma \, \Tilde{m}^{-2}} \label{eq_m_small_PL}   \\
\lim_{m_{\gamma^\prime} \rightarrow \infty} \chi^2 & = & 2 \Tilde{m} R \label{eq_m_large_PL} \, ,
\end{eqnarray}
\end{subequations}
meaning that, for large masses $\chi \sim \sqrt{m_{\gamma^\prime}}$, whereas for small masses $\chi \sim 1/m_{\gamma^\prime}$ and the renormalization procedure suggested in Ref.~\cite{Joerg_spectrum} and performed in the previous section is not necessary. The reason for this is simple: our observable, the sensitivity $\gamma$, does not explicitly depend on electromagnetic parameters, which drop out from the ratio of $V(R^*)$ and $V(R)$, cf.~Eq.~\eqref{eq_gamma}.

\begin{figure}[t!]
\begin{minipage}[b]{1.\linewidth}
\includegraphics[width=\textwidth]{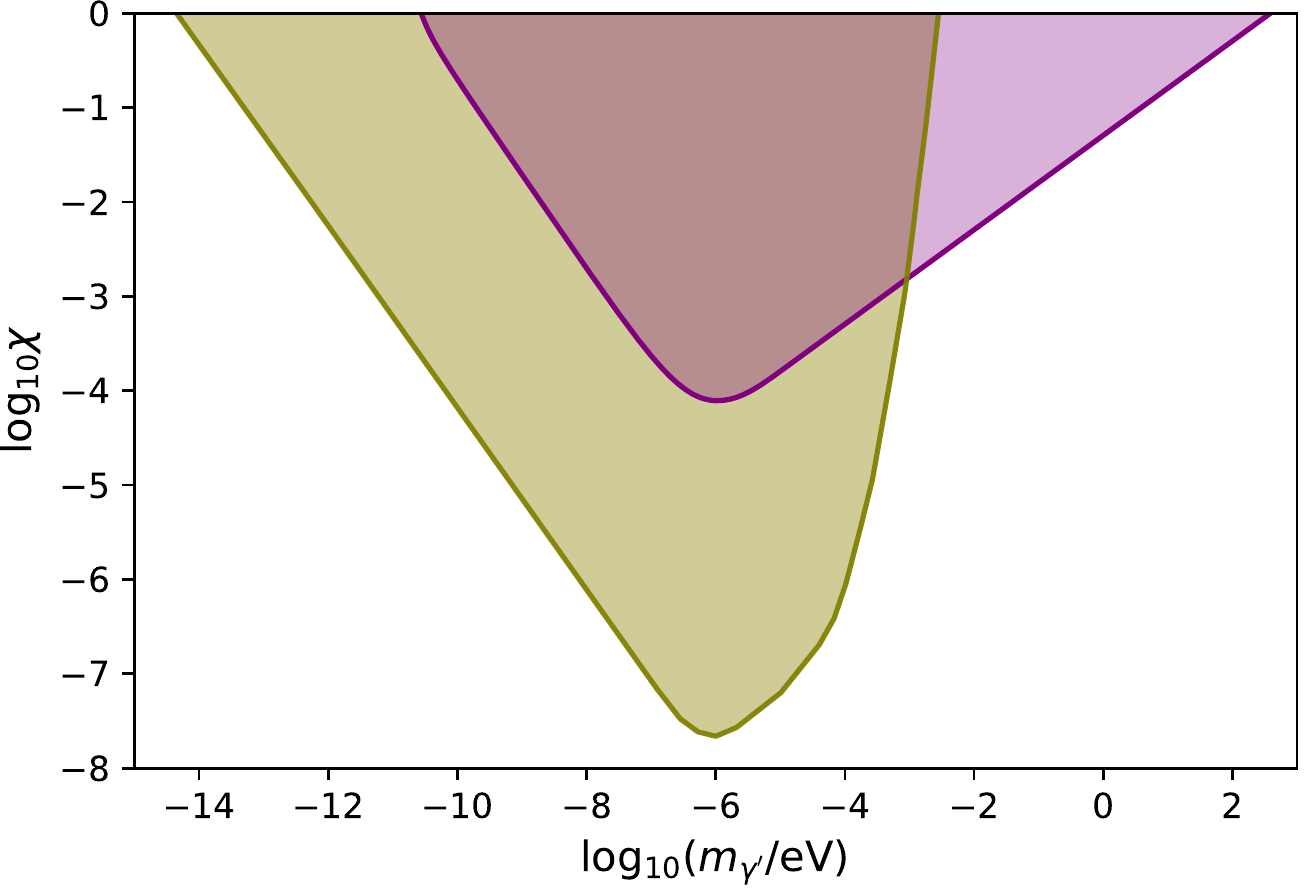}
\end{minipage} \hfill
\caption{Excluded region in parameter space for hidden photons based on Cavendish-like~\cite{JJ, Tu_2005, Williams, Bartlett} (green) and Plimpton-Lawton experiments~\cite{Plimpton} (purple), cf.~Eq.~\eqref{eq_limit_plimpton}.}
\label{fig_plimpton}
\end{figure}

In the Plimpton-Lawton experiment, the accuracy of the galvanometer was $\sim 1 \, \mu$V and no potential difference between the metal shells was detected. Using $V_0 = 3$~kV we have a sensitivity $\gamma = 3.3 \times 10^{-10}$ so that, with Eq.~\eqref{eq_limit_plimpton} we are able to exclude the purple region in Fig.~\ref{fig_plimpton}, in which we also include the exclusion limits from Cavendish-like experiments~\cite{JJ, Tu_2005, Williams, Bartlett}.

The qualitative features of our limit deserve some remarks. In Ref.~\cite{Bartlett} the authors review the experiments performed in the late 1960 and early 1970, which employed three to four metallic shells held at different potentials. The general strategy was to look for unexpected potential differences between the inner shells once the exterior ones are held at some known potential. Let us consider an experiment with four concentric shells of radii $a < b < c < d$ in which the potential difference $V_{cd}$ is held fixed by the experimenters. The potential difference between the inner shells $V_{ab}$ should be zero in a perfectly Coulombian world and a deviation from this could be a potential signal of new physics. The sensitivity in this situation is $\gamma^* = V_{ab}/V_{cd}$, which is typically better than one part in $10^{12}$.

%
%

For small masses the bound from the multi-shell Cavendish-like experiments displays the asymptotic behavior shown in Eq.~\eqref{eq_m_small_PL}. For $\Tilde{m}^2 \lesssim 6\gamma/(R^2 - R^{*2})$, the bound rises much more sharply -- with the physical parameters of the Plimpton-Lawton setup this happens at a mass $m_{\gamma^\prime} \approx 0.02$~neV. Interestingly enough, in the large-mass limit the Cavendish-like and Plimpton-Lawton setups provide very different bounds. This is due to the presence of an exponential factor $\chi^2 \sim \exp{\left(+\Tilde{m}(c-b)\right)}$~\cite{Bartlett} that is not present in the Plimpton-Lawton setup, cf. Eq.~\eqref{eq_m_large_PL}. Besides this, three radii, namely $b,c$ and $d$, contribute in this limit and it is clear that this result cannot be easily reduced to the two-shell configuration of the Plimpton-Lawton experiment. The reason for this are the fundamental differences in what is actually being measured in each experiment.

To see this, let us look at the setups in more detail. In the Plimpton-Lawton experiment there are only two shells -- the outer one held at a potential -- and the sensitivity is simply the potential difference between them. In this case, there are only two radii that can appear in the sensitivity, cf. Eqs.~\eqref{eq_h} and~\eqref{eq_g}. The Cavendish-like setups are more complex: the two outer shells are held at a given potential and the signal to be measured is the potential difference between the two innermost shells. This means that there is no direct measurement of the potential difference between inner and outer shells. Due to the functional form of the factors appearing in our Eqs.~\eqref{eq_h},~\eqref{eq_g} and in Eq.~(6) from Ref.~\cite{Bartlett}, we see that the small-mass limit must be the same. However, the fact that the sensitivity in the Cavendish-like experiments does not include $V_{bc}$ -- or any combination of inner and outer shells, such as $V_{bd}$, $V_{ac}$ or $V_{ad}$ -- makes a sensible comparison with our limits impossible, i.e, not only the geometries are different, but also the quantities being measured. This is particularly visible in the large-mass limit.


\begin{figure}[t!]
\begin{minipage}[b]{1.\linewidth}
\includegraphics[width=\textwidth]{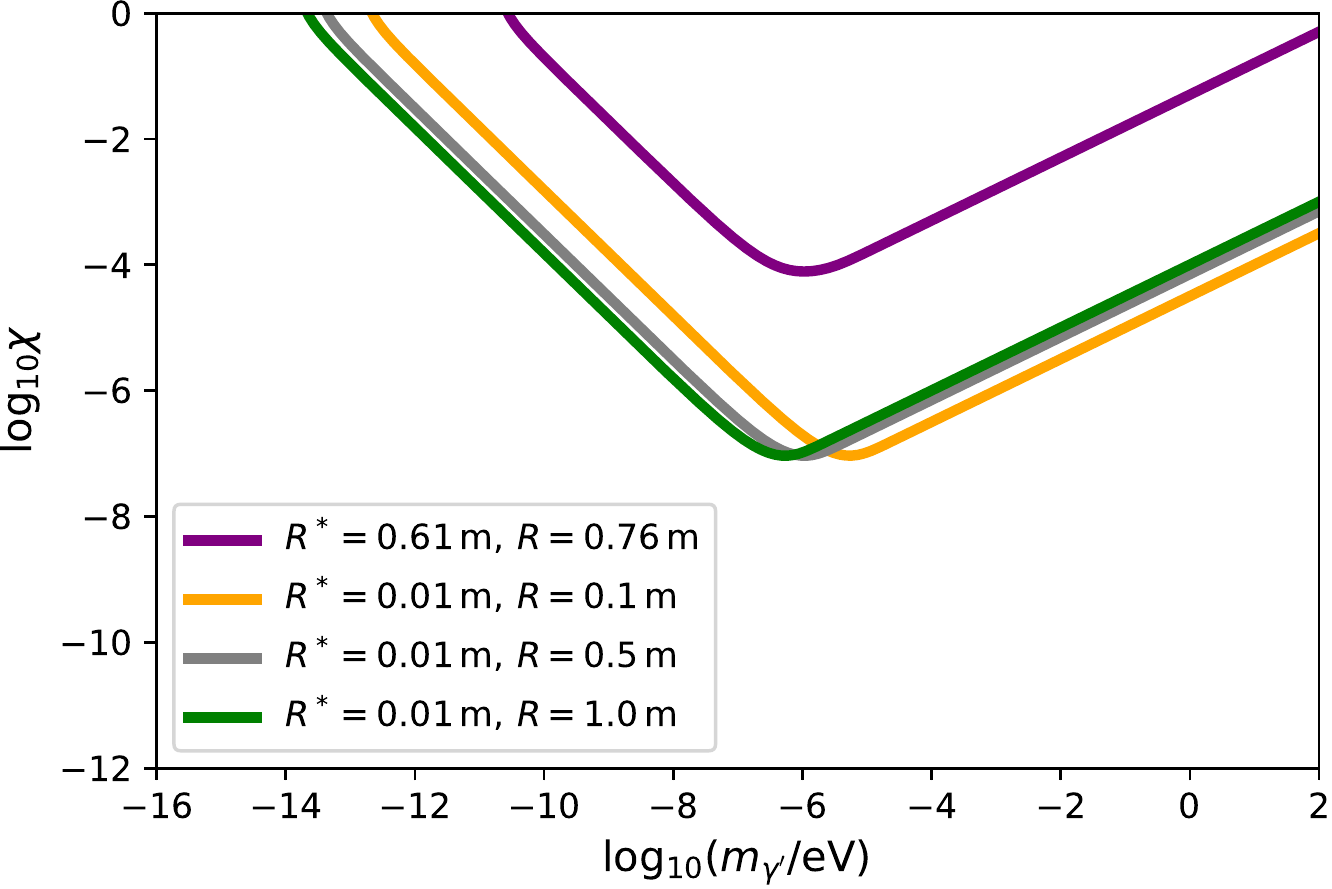}
\end{minipage} \hfill
\caption{Bounds on hidden photons based on the Plimpton-Lawton experiment~\cite{Plimpton}. In purple we show the bound with the original parameters. The other curves show the projected limits if we consider spheres with different radii and assume a sensitivity $\gamma = 10^{-15}$, compatible with the one reached in more recent Cavendish-like experiments~\cite{Tu_2005}.}
\label{fig_plimpton_alt}
\end{figure}

Plimpton and Lawton performed their experiment almost 90 years ago and, even with a worse sensitivity and a simpler arrangement, their setup provides a limit that goes beyond the excluded region from Cavendish-like experiments for $m_{\gamma^\prime} \gtrsim $~meV. It is worthwhile checking whether the bounds could be improved if a similar setup would be designed today. In particular, we wish to find an optimal set $(R^*, R)$ that would increase the area of no overlap to the right on Fig.~\ref{fig_plimpton}; i.e., the area which is originally constrained only by astrophysical tests (cf. gray region in Fig.~\ref{fig_exclusion_HP}). For small masses, our bounds behave according to Eq.~\eqref{eq_m_small_PL} and we see that $\chi \sim 1/(m_{\gamma^\prime}\sqrt{R^2 - R^{*2}})$. Therefore, it is advantageous to maximize the denominator by choosing as small $R^*$ and large $R$ as practically possible. On the other hand, for large masses, the inner radius only appears in $h(m_{\gamma^\prime})$, cf.~Eq.~\eqref{eq_h}, but this function tends to zero in the large-mass limit. Consequently, only $g(m_{\gamma^\prime})$, cf.~Eq.~\eqref{eq_g}, contributes, and the value of $R^*$ is irrelevant. With this, the bound behaves as $\chi \sim \sqrt{m_{\gamma^\prime} R}$ and it is better to choose a small $R$.

In the intermediate mass range the radius of the outer sphere sets the scale for the highest sensitivity. Since we are interested in increasing the non-overlap with the excluded region from the Cavendish-like experiments (cf.~Fig.~\ref{fig_plimpton}), we should move our bound as much towards large masses as possible. This means that we should choose a radius $R < 76$~cm, but a drastic reduction may not be experimentally feasible. For this reason we fix $R
^*$ at 1~cm and vary $R$ from 10 to 100~cm, as shown in Fig.~\ref{fig_plimpton_alt}. The most interesting choice is then clearly the orange curve, which is also shown in Fig.~\ref{fig_exclusion_HP} together with other bounds.


\section{Concluding remarks} \label{sec_conclusions}

\indent

We have considered modifications to classical electrodynamics caused by the presence of hidden photons. These hypothetical spin-1 bosons would be kinetically mixed with the photon and could generate small modifications to Coulomb's law that could be detected in sensitive experiments. With this in mind, we have analysed two AFM setups~\cite{Mohideen1, Mohideen2} and the Plimpton-Lawton experiment~\cite{Plimpton}. The former were originally conceived to measure the Casimir force at sub-$\mu$m distances, whereas the latter is a null test of deviations from Coulomb's law at the $\sim 10$~cm scale. In both cases, hidden photons could be detected via Yukawa-like terms in the force~\eqref{eq_force} or in the potential~\eqref{eq_pot_plimpton}, respectively.


In our analyses, we have excluded regions in parameter space which were mostly already constrained by other laboratory-based tests, such as Cavendish-like experiments and atomic spectroscopy, therefore reinforcing those limits. The gray regions in Fig.~\ref{fig_exclusion_HP} are constrained only by astrophysical and cosmological observations, which, despite being very sensitive, are more model dependent than table-top, laboratory experiments such as the ones considered here. Let us, then, briefly discuss how to improve our limits.

Our bound from the AFM experiments is shown in red in Figs.~\ref{fig_exclusion_HP} and~\ref{fig_exclusion_diff_a}, while the black curve displays a projection if other sphere-plane separations are considered with a 2-fold reduction in the experimental uncertainty. Nonetheless, the latter still overlaps with the region excluded in Ref.~\cite{Joerg_spectrum}, which starts at $m_{\gamma^\prime} \simeq 4$~meV. It is, therefore, important to consider measurements with even larger sphere-plane separations -- reaching ideally a few $\mu$m -- in order to move the bound further towards lower masses. More recent AFM measurements of the gradient of the Casimir force, employing the same geometry as the one analysed here, have been conducted over $\mu$m-scale sphere-plane separations by the same group~\cite{Mohideen5, Mohideen6}. For separations close to 200~nm they do reach a better relative uncertainty of $\sim 0.3\%$ by reducing the residual electrostatic potential through better cleaning techniques. At this distance, however, we do not expect a large enough improvement on the red curve in Fig.~\ref{fig_exclusion_diff_a}. Moreover, the relative experimental uncertainties attained at the largest -- and more interesting -- separations around $\mu m$ are at least 15 times larger than those quoted in Refs.~\cite{Mohideen1, Mohideen2}, cf.~Table~\ref{table_AFM}. This would dramatically reduce the already limited reach of our AFM bounds, so we refrain from pursuing a more detailed analysis.


The Plimpton-Lawton experiment yields bounds that go beyond the limits from Cavendish-like tests; thus, covering regions previously tested only by astrophysical observations, cf.~Fig.~\ref{fig_plimpton}. This is a welcome result, since it strengthens our confidence in the exclusion of those areas in parameter space. It is worth keeping in mind that the original experiment was performed in 1936 and, since then, the instrumentation and experimental techniques have drastically improved. Already in 1971 Williams {\it et al.}~\cite{Williams} used a lock-in to measure potential differences at the pV level, many orders of magnitude better than what Plimpton and Lawton could achieve. A similarly high sensitivity, $\gamma = 10^{-15}$, was assumed to obtain the projected bounds shown in Fig.~\ref{fig_plimpton_alt}. Furthermore, as already discussed, it is important to reduce the radii as much as practically possible, so as to shift the bound toward larger hidden photon masses. These strategies go in the opposite direction of what is typically needed to set stronger upper bounds on the photon mass~\cite{Tu_2005}. In any case, higher applied voltages and better sensitivities to potential differences could be used to surpass even the most optimistic projection in Fig.~\ref{fig_plimpton_alt}, potentially crossing into unconstrained regions in the parameter space.

As a closing remark, let us mention that the experiments considered here were designed to probe features of standard electrodynamics: classical in the case of the Plimpton-Lawton experiment and quantum-mechanical in the AFM measurements. Nonetheless, the respective results could be reinterpreted to constrain hidden photons, i.e., new physics. In this sense, an improved redesign of e.g. the Plimpton-Lawton setup could be performed not (only) to search for an even better lower bound on the photon mass, but rather to indirectly detect new physics, be it massive hidden photons or other beyond the SM particle that may alter the electromagnetic properties of the system. Therefore, it would be worthwhile revisiting the roughly 90-year-old experiment of Plimpton and Lawton to tackle more contemporary questions.




\section*{Acknowledgements}

\indent 

We acknowledge correspondence with U. Mohideen, A. Caputo and P. Fayet. The authors are grateful to J. A. Helayël-Neto and J. Jaeckel for helpful comments and for reading the manuscript. We also thank the anonymous referee for his/her relevant comments.


\begin{thebibliography}{99}




\bibitem{Bilenky} S. Bilenky, Neutrino oscillations: From a historical perspective to the present status, Nucl. Phys. B \textbf{908}, 2 (2016).

\bibitem{DM} K. Arun, S.B. Gudennavar, C. Sivaram, Dark matter, dark energy, and alternate models: A review, Advances in Space Research {\bf 60}, 166 (2017).

\bibitem{Ellis} J. Elis, Outstanding questions: physics beyond the Standard Model, Phil. Trans. Roy. Soc. Lond. A {\bf 370}, 818 (2012).

\bibitem{Kasakov} D.I. Kasakov, Prospects of elementary particle physics, Physics Uspekhi {\bf 62}, 4 364 (2019).

\bibitem{Okun} L.B. Okun, Limits of electrodynamics: paraphotons?, Sov. Phys. JETP {\bf 56}, 502 (1982).

\bibitem{Holdon} B. Holdom, Two U(1)'s and epsilon charge shifts, Phys. Lett. B {\bf 166}, 196 (1986).

\bibitem{Abel1} S.A. Abel, M.D. Goodsell, J. Jaeckel, V.V. Khoze, A. Ringwald, Kinetic mixing of the photon with hidden U(1)s in string phenomenology, JHEP {\bf 807}, 124 (2008).

\bibitem{Abel2} S. Abel, J. Santiago, Constraining the string scale: from Planck to Weak and back again, J. Phys. G {\bf 30}, (2004).

\bibitem{Fayet1} P. Fayet, Extra U(1)'s and new forces, Nucl. Phys. B {\bf 347}, 743 (1990).

\bibitem{JJ} J. Jaeckel, A force beyond the Standard Model - Status of the quest for hidden photons, arXiv:hep-ph/1303.1821.

\bibitem{Fabbrichesi} M. Fabbrichesi, E. Gabrielli, G. Lanfranchi, The dark phhoton, arXiv:hep-ph/2005.01515.

\bibitem{Mohideen1} A. Roy, C.-Y. Lin, U. Mohideen, Improved precision measurement of the Casimir force, Phys. Rev. D {\bf 60}, 111101(R) (1999).

\bibitem{Mohideen2} B.W. Harris, F. Chen, U. Mohideen, Precision measurement of the Casimir force using gold surfaces, Phys. Rev. A {\bf 62}, 052109 (2000).

\bibitem{Plimpton} S. Plimpton, W. Lawton, {\it A very accurate test of Coulomb's law of force between charges}, Phys. Rev. {\bf 50}, 1066 (1936). 

\bibitem{Mohideen3} U. Mohideen, A. Roy, Precision measurement of the Casimir force from 0.1 to 0.9 $\mu$m, Phys. Rev. Lett. {\bf 81}, 4549 (1998).

\bibitem{Mohideen4} A. Roy, U. Mohideen, Demonstration of the nontrivial boundary dependence of the Casimir force, Phys. Rev. Lett. {\bf 82}, 4380 (1999).

\bibitem{Caputo1} A. Caputo, H. Liu, S. Mishra-Sharma, J.T. Ruderman, Dark photon oscillations in our inhomogeneous universe, arXiv:astro-ph.CO/2002.05165.

\bibitem{Caputo2} A. Caputo, H. Liu, S. Mishra-Sharma, J.T. Ruderman, Modeling dark photon oscillations in our inhomogeneous universe, arXiv:astro-ph.CO/2004.06733.

\bibitem{Tu_2005} Liang-Cheng Tu, Jun Luo, George T Gillies, The mass of the photon, Rept. Prog. Phys. {\bf 68}, 77 (2005).

\bibitem{Williams} E.R. Williams, J.E. Faller H.A. Hill, New experimental test of Coulomb’s law: A laboratory upper limit on the photon rest mass, Phys. Rev. Lett. {\bf 26}, 721 (1971).

\bibitem{Bartlett} D.F. Bartlett, S. Loegl, Limits on an electromagnetic fifth force, Phys. Rev. Lett. {\bf 61}, 2285 (1988).

\bibitem{Joerg_spectrum} J. Jaeckel, S. Roy, Spectroscopy as a test of Coulomb’s law: A probe of the hidden sector, Phys. Rev. D {\bf 82}, 125020 (2010).

\bibitem{Parthey} C.G. Parthey {\it et al.}, Improved measurement of the hydrogen $1s-2s$ transition frequency, Phys. Rev. Lett. {\bf 107}, 203001 (2011).

\bibitem{Goldhaber1} A.S. Goldhaber, NM.M. Nieto, Terrestrial and extraterrestrial limits on the photon mass, Rev. Mod. Phys. {\bf 43}, 277 (1971).

\bibitem{Goldhaber2} A.S. Goldhaber, NM.M. Nieto, New geomagnetic limit on the mass of the photon, Phys. Rev. Lett. {\bf 21}, 567 (1968).

\bibitem{Ringwald} J. Jaeckel, A. Ringwald, The low-energy frontier of particle physics, Annual Review of Nuclear and Particle Science {\bf 60}, 405 (2010).










\bibitem{Holstein1} B.R. Holstein, A. Ross, {\it Spin effects in long range electromagnetic scattering}, arXiv:hep-ph/0802.0715.

\bibitem{Holstein2} B.R. Holstein, {\it Analytical on-shell calculation of low energy higher order scattering}, J. Phys. G {\bf 44}, 1 (2017).

\bibitem{pots_HP_cbpf} G.P. de Brito, P.C. Malta, L.P.R. Ospedal, Spin- and velocity-dependent nonrelativistic potentials in modified electrodynamics, Phys. Rev. D \textbf{95}, 016006 (2017).



















\bibitem{Casimir1} H.B.G. Casimir, On the attraction between two perfectly conducting plates, Proceedings of the Royal Netherlands Academy of Arts and Sciences {\bf 51}, 793 (1948).

\bibitem{Lif} E.M. Lifshitz, The theory of molecular attractive forces between solids, Sov. Phys. JETP {\bf 2}, 73 (1956).

\bibitem{Casimir2} S.K. Lamoreaux, Demonstration of the Casimir force in the 0.6 to 6 $\mu$m range, Phys. Rev. Lett. {\bf 78}, 5 (1997).

\bibitem{Casimir3} P.H.G.M. van Blokland, J.T.G. Overbeek, van der Waals forces between objects covered with a chrome layer, J. Chem. Soc. Faraday Trans. I {\bf 74}, 2651 (1978).

\bibitem{Casimir4} G. Bressi, G. Carugno, R. Onofrio, G. Ruoso, Measurement of the Casimir force between parallel metallic surfaces, Phys. Rev. Lett. {\bf 88}, 041804 (2002).

\bibitem{Blocki} J. Blocki, J. Randrup, W.J. Swiatecki, C.F. Tsang, Proximity forces, Ann. Phys. {\bf 105}, 427 (1977).

\bibitem{Mazur} A.A. Maradudin, P. Mazur, Effects of surface roughness on the van der Waals force between macroscopic bodies, Phys. Rev. B {\bf 22}, 1677 (1980). 

\bibitem{Novello1} V.M. Mostepanenko, M. Novello, Constraints on non-Newtonian gravity from the Casimir force measurements between two crossed cylinders, Phys. Rev. D {\bf 63}, 115003 (2001).

\bibitem{Novello2} E. Fischbach, D.E. Krause, V.M. Mostepanenko, M. Novello, New constraints on ultrashort-ranged Yukawa interactions from atomic force microscopy, Phys. Rev. D {\bf 64}, 075010 (2001).

\bibitem{Bordag1} M. Bordag, B. Geyer, G.L. Klimchitskaya, V.M. Mostepanenko, Constraints for hypothetical interactions from a recent demonstration of the Casimir force and some possible improvements, Phys. Rev. D {\bf 58}, 075003 (1998).

\bibitem{Bordag2} M. Bordag, B. Geyer, G.L. Klimchitskaya, V.M. Mostepanenko, New constraints for non-Newtonian gravity in the nanometer range from the improved precision measurement of the Casimir force, Phys. Rev. D {\bf 62}, 011701 (2000).

\bibitem{Bordag3} M. Bordag, B. Geyer, G.L. Klimchitskaya, V.M. Mostepanenko, Stronger constraints for nanometer scale Yukawa-type hypothetical interactions from the new measurement of the Casimir force, Phys. Rev. D {\bf 60}, 055004 (1999).


\bibitem{jackson1999} J.~D.~Jackson, \emph{Classical Electrodynamics} (John Wiley and Sons, New York, 1999).

\bibitem{griffiths1999} D.~J.~Griffiths, \emph{Introduction to Electrodynamics} (Prentice Hall, Upper Saddle River, New Jersey, 1999)

\bibitem{Morrison89} C.~A.~Morrison, The Potential and Electric Fields of a Conducting Sphere in the Presence of a Charged Conducting Plane, HDL-TR-2161 (1989).

\bibitem{Dallagnol2009} F.~F.~Dall'Agnol, V.~P.~Mammana, Solution for the electric potential distribution produced by sphere-plane electrodes using the method of images, Rev. Bras. Ensino Fís. {\bf 31} 3503.1 (2009).























\bibitem{Mohideen5} M. Liu, J. Xu, G.L. Klimchitskaya, V.M. Mostepanenko, U. Mohideen, Examining the Casimir puzzle with upgraded technique and advanced surface cleaning, Phys. Rev. B {\bf 100}, 081406(R) (2019).

\bibitem{Mohideen6} M. Liu, J. Xu, G.L. Klimchitskaya, V.M. Mostepanenko, U. Mohideen, Precision measurements of the gradient of the Casimir force between ultra clean metallic surfaces at larger separations, Phys. Rev. A {\bf 100}, 052511 (2019).





















\end{thebibliography}
\end{document}